\documentclass[pdflatex,sn-standardnature]{sn-jnl}
\jyear{2025}
\usepackage{setspace}
\usepackage{lineno}
\usepackage{graphicx}
\usepackage{multirow}
\usepackage{amsmath,amssymb,amsfonts}
\usepackage{amsthm}
\usepackage{mathrsfs}
\usepackage[title]{appendix}
\usepackage{xcolor}
\usepackage{textcomp}
\usepackage{manyfoot}
\usepackage{booktabs}
\usepackage{algorithm}
\usepackage{algorithmicx}
\usepackage{algpseudocode}
\usepackage{listings}
\usepackage{textcomp}
\usepackage{subcaption}
\usepackage{array}
\usepackage{threeparttable}

\begin{document}

\title[Article Title]{CondensNet: Enabling stable long-term climate simulations via hybrid deep learning models with adaptive physical constraints}


\author[1]{\fnm{Xin} \sur{Wang}}\email{xin.w24@nus.edu.sg}

\author[2]{\fnm{Jianda} \sur{Chen}}\email{chenjd21@mails.tsinghua.edu.cn}

\author[3]{\fnm{Juntao} \sur{Yang}}\email{yjuntao@nvidia.com}

\author[3]{\fnm{Jeff} \sur{Adie}}\email{jadie@nvidia.com}

\author[3]{\fnm{Simon} \sur{See}}\email{ssee@nvidia.com}

\author[4]{\fnm{Kalli} \sur{Furtado}}\email{kalli\_furtado@nea.gov.sg}

\author[4]{\fnm{Chen} \sur{Chen}}\email{chen\_chen@nea.gov.sg}

\author[5]{\fnm{Troy} \sur{Arcomano}}\email{tarcomano@anl.gov}

\author[6]{\fnm{Romit} \sur{Maulik}}\email{rmaulik@psu.edu}

\author[2]{\fnm{Wei} \sur{Xue}}\email{xuewei@tsinghua.edu.cn}

\author*[1,7]{\fnm{Gianmarco} \sur{Mengaldo}}\email{mpegim@nus.edu.sg}

\affil*[1]{
\orgname{Department of Mechanical Engineering, National University of Singapore}, 
\country{SG}}

\affil[2]{
\orgdiv{Department of Computer Science and Technology, Tsinghua University}, \country{CN}}


\affil[3]{
\orgdiv{NVIDIA AI Technology Centre}, 
\orgname{NVIDIA Corporation}, 
\country{SG}}

\affil[4]{
\orgdiv{Centre for Climate Research Singapore}, 
\country{SG}}

\affil[5]{
\orgdiv{Environmental Science Division, Argonne National Laboratory}, 
\country{USA}}

\affil[6]{
\orgdiv{Information Sciences and Technology Department, The Pennsylvania State University}, 
\country{USA}}

\affil[7]{
\orgdiv{Department of Mathematics, National University of Singapore \\(by courtesy)}, \country{SG}}


\abstract{Accurate and efficient climate simulations are crucial for understanding Earth's evolving climate. 
However, current general circulation models (GCMs) face challenges in capturing unresolved physical processes, such as cloud and convection. 
A common solution  is to adopt cloud resolving models, that provide more accurate results than the standard subgrid parametrization schemes typically used in GCMs. 
However, cloud resolving models, also referred to as super parametrizations, remain computationally prohibitive. 
Hybrid modeling, which integrates deep learning with equation-based GCMs, offers a promising alternative but often struggles with long-term stability and accuracy issues.
In this work, we find that water vapor oversaturation during condensation is a key factor compromising the stability of hybrid models. 
To address this, we introduce CondensNet, a novel neural network architecture that embeds a self-adaptive physical constraint to correct unphysical condensation processes. 
CondensNet effectively mitigates water vapor oversaturation, enhancing simulation stability while maintaining accuracy and improving computational efficiency compared to super parametrization schemes.
We integrate CondensNet into a GCM to form PCNN-GCM (Physics-Constrained Neural Network GCM), a hybrid deep learning framework designed for long-term stable climate simulations in real-world conditions, including ocean and land. 
PCNN-GCM represents a significant milestone in hybrid climate modeling, as it shows a novel way to incorporate physical constraints adaptively, paving the way for accurate, lightweight, and stable long-term climate simulations.
}

\maketitle

\section{Introduction}
\label{sec:introduction}

Climate change is bringing more frequent and intense extreme weather events that are causing significant harm to ecosystems and communities worldwide~\cite{zhongming2021ar6}. 

General circulation models (GCMs), climate models that use mathematical equations to simulate the Earth's system, are critical to understand climate change over various time scales~\cite{allan2021ipcc}. 
Indeed, they can be pivotal for the survival of a certain community or ecosystem, if they were to provide reliable projections on actionable time scales. 

One of the primary sources of unreliability in modern GCMs stems from the challenges associated with \textit{cloud} and \textit{convection} processes. 
These phenomena are inherently tied to small-scale atmospheric physics occurring at kilometer or sub-kilometer scales, which current GCMs struggle to capture due to their relatively coarse spatial resolution ($\sim$ 50 km). 
To address this limitation, various approaches have been proposed in the literature.
These can be grouped into two categories: subgrid parametrization models, and super parametrization models. 

Subgrid parametrization models use simplified empirical relationships or theoretical approximations to describe small-scale atmospheric processes that current GCMs are unable to resolve~\cite{bony2015clouds,liang2022stiffness,randall2003breaking,emanuel1994large}. 
While computationally efficient, these subgrid models are relatively simplistic, and are a leading source of uncertainties in GCMs' climate projections~\cite{stevens2013climate}.

Super parametrization models embed high-resolution cloud-resolving models (CRMs) within each grid cell of a larger-scale GCMs.
These CRMs explicitly resolve small-scale atmospheric processes, enhancing the model's ability to simulate cloud dynamics. 
A notable example is the super-parameterized community atmosphere model (SPCAM), developed by the National Center for Atmospheric Research (NCAR)~\cite{khairoutdinov2001cloud,khairoutdinov2005simulations}. 
Despite their improved accuracy, these models are computationally intensive, often rendering them impractical for long-term climate projections~\cite{satoh2008nonhydrostatic}.

A promising and emerging area of research, known as hybrid modeling (as it hybridizes physics and machine learning), blends CRMs with machine learning.
The idea is to use the wealth of scale-resolving data from various CRMs to construct high-fidelity and computationally-efficient deep learning (DL) emulators (also referred to as DL parametrizations) of small-scale processes, such as cloud and convection processes, on each cell of the coarse GCM grid. 
This promising research area offers better accuracy than traditional subgrid parametrization models at a much cheaper computational cost than super parametrization models~\cite{rasp2018deep,yuval2020stable,han2020moist,mooers2021assessing,arcomano2022hybrid}. 

While several breakthroughs have been made in hybrid climate modeling over the past few years, one key issue remains: the lack of stability for long-term climate simulations.

To address this issue, many attempts focused on enforcing physical constraints within machine learning models, initially on idealized settings~\cite{beucler2019achieving,beucler2021enforcing,yuval2020stable,yuval2021use}. 
These efforts demonstrated some degree of success, without fully addressing the lack of long-term simulation stability, especially in real-world settings (a real-world setting is a present-day climate model setting, coupled to a land surface model Community Land Model version 4.0~\cite{oleson2010technical} and forced under prescribed sea surface temperatures and sea ice concentrations~\cite{hurrell2008new}).  
More recently, 10-year stable simulations under real land and sea distributions were achieved~\cite{wang2022stable}, although with a trial-and-error approach, and without fully addressing the issue. 
Recent work has targeted specific aspects, such as relative humidity to prevent excessive moistening~\cite{fuchs2023torchclim} and water condensation, enabling 5-month stable runs~\cite{behrens2024improving}.
Physical constraints like global mass and moisture conservation were incorporated into ACE2, a promising DL-based climate emulator with a post-processing correction module~\cite{watt2024ace2}.
Another promising approach used differentiable dynamics to integrate neural networks with numerical methods, improving stability and prediction~\cite{kochkov2024neural}, without directly enforcing physical constraints. 
Recent results also showed stable simulations for hybrid models with single-module DL parametrizations (i.e., convection-only; radiation-only)~\cite{balogh2025online,hafner2025stable}.

\begin{figure}[htbp]
  \centering
    \includegraphics[width=1.00\linewidth]{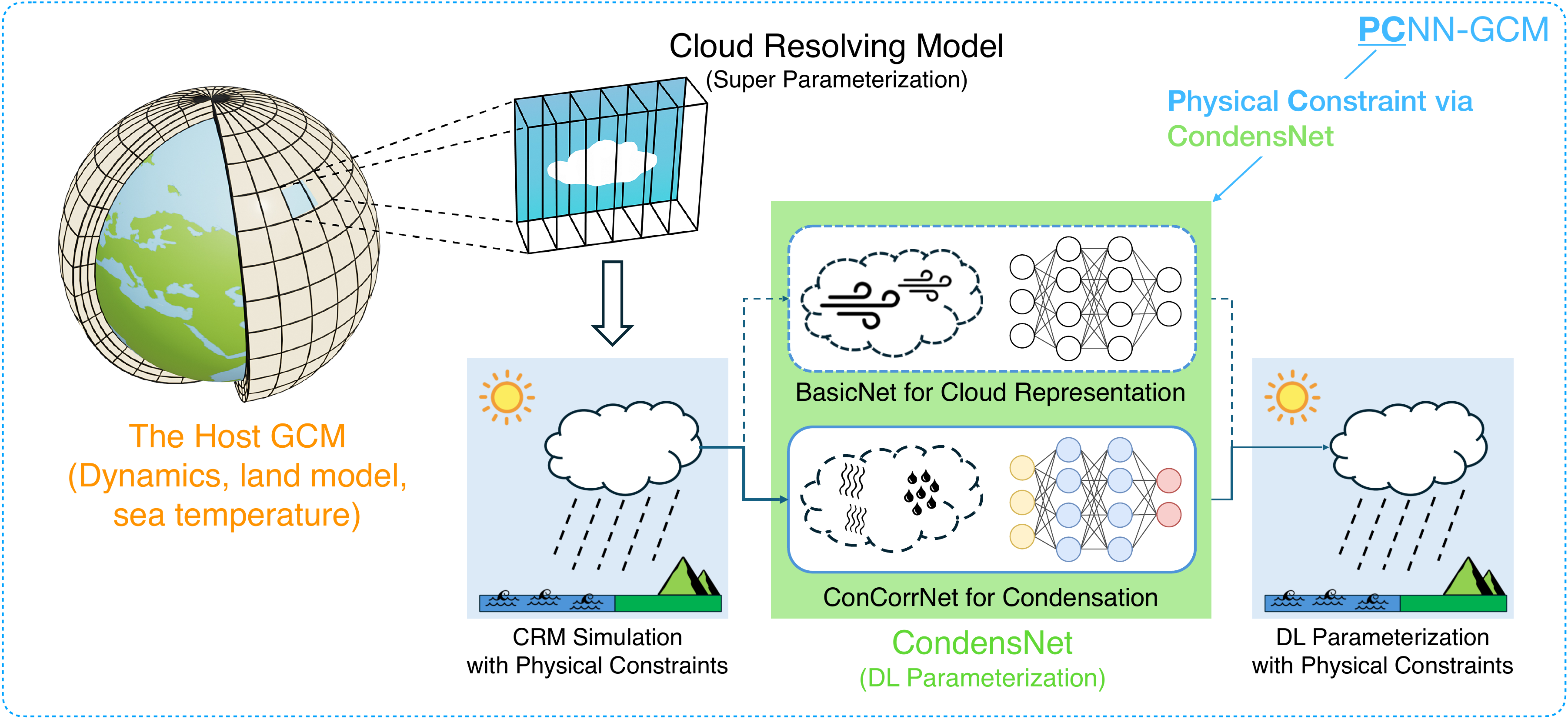}
    \caption{Methodology of the CondensNet model. CondensNet is a physically-constrained DL parametrization coupled with a climate dynamics engine to support hybrid modeling. The network architecture mainly has two parts: BasicNet for learning the cloud representation and ConCorrNet for condensation physical constraint.}
  \label{fig:Fig1}
\end{figure}

In this study, we show that the regulation of water vapor saturation is critical to obtain long-term stability in hybrid DL climate models, and to avoid oversaturation that in turn leads the simulations to fail.
To address this problem, we propose a new neural network architecture, depicted in Figure~\ref{fig:Fig1}, constituted of two main components: 1) a basic model (BasicNet), learning the cloud representation, and 2) a condensation correction network (ConCorrNet), learning the condensation process. 
We name this new neural network architecture, composed of the BasicNet and the ConCorrNet, ~\textit{CondensNet}, to emphasize the physical condensation constraint imposed. 
Here, by \textit{cloud representation} we refer to the grid-scale parametrization of subgrid cloud and convective processes: the mapping from large-scale prognostic states and forcings to column tendencies at the model resolution. Concretely, in this work it means predicting the water vapor and dry-static-energy tendencies ($\mathrm{d}Q$ and $\mathrm{d}s$) at each vertical level, while radiative effects are produced by a separate DL radiation emulator and are not modified by the condensation constraint.
Notably, unlike simple post‐processing methods that mechanically adjust non‐physical predictions solely to enforce water vapor conservation (often resulting in inaccurate solutions -- e.g., significant water vapor biases), our ConCorrNet adaptively refines BasicNet outputs by learning from SPCAM training targets, thereby ensuring that corrections are both physically consistent and accurate, in contrast to those obtained from an unconstrained neural network.
With a particular training strategy (detailed in Methods, section~\ref{sec:methods-training}), CondensNet can accurately learn cloud physics from super parametrization (CRM) simulations, while imposing a physical constraint that address the water vapor oversaturation issue. 
CondensNet is integrated with the Community Atmosphere Model (CAM), that is used as our reference host GCM. 
CAM is responsible to drive the large-scale dynamics of the hybrid climate simulation, while CondensNet learns small-scale processes from SPCAM.
The latter is our reference super parametrization, also referred to as SP-GCM to highlight that CondensNet can be applied to any super parametrization scheme and host GCM. 

CondensNet together with the chosen host GCM, namely CAM, form the overall hybrid DL-GCM framework that we name \textit{PCNN-GCM} (\textit{Physics Constrained-Neural Network GCM}).
PCNN-GCM runs under real-world conditions, as it uses Community Land Model version 4.0~\cite{lawrence2011parametrization} for the land component of the Earth system, and it is forced under prescribed sea surface temperatures and sea ice concentrations according to the Atmospheric Model Intercomparison Project (AMIP) protocol. 
The new PCNN-GCM framework represents a major step forward in hybrid climate modeling, as it showcases the effective use of adaptive physical constraints to achieve long-term stability without sacrificing accuracy, at a much cheaper computational cost than super parametrization models.
PCNN-GCM additionally addresses one of the key drawbacks in previous stable simulations, namely the need for extensive and time-consuming trial-and-error before achieving stability~\cite{wang2022stable}, or the reliance on online training and extensive software engineering resources for developing fully differentiable dynamics~\cite{kochkov2024neural}.

We note that PCNN-GCM is also faster and more accurate, in some respects (such as precipitation and intraseasonal variability), than CAM5, and (unlike CAM) capable of significant GPU acceleration even on a modest number of nodes. 
For instance, on one 24-core node, PCNN-GCM is nearly 4 times faster than CAM5 when accelerated on GPU -- see section~\ref{sec:computational-costs}.

We remark that CondensNet is coupled to the CAM/CESM platform; the lightweight Fortran–Python Interface (see Supplementary Information section~D) is framework-agnostic and provides a standardized physics–parametrization interface that aligns with how many GCMs exchange model states and tendencies. 
After model-specific variable and vertical mapping, DL parametrizations can be iterated with minimal host-side engineering, enabling researchers to focus on algorithmic innovation, hence the naming choice, i.e., PCNN-GCM.

\section{Results}
\label{sec:results}
%
\subsection{Long-term stability}
\label{subsec:stablity}
GCMs integrated with DL parametrizations often become unstable during long-term simulations, where the instability typically manifests as an energy surge~\cite{wang2022stable}.

\begin{figure}[htbp]
  \centering        
    \includegraphics[width=1.00\linewidth]{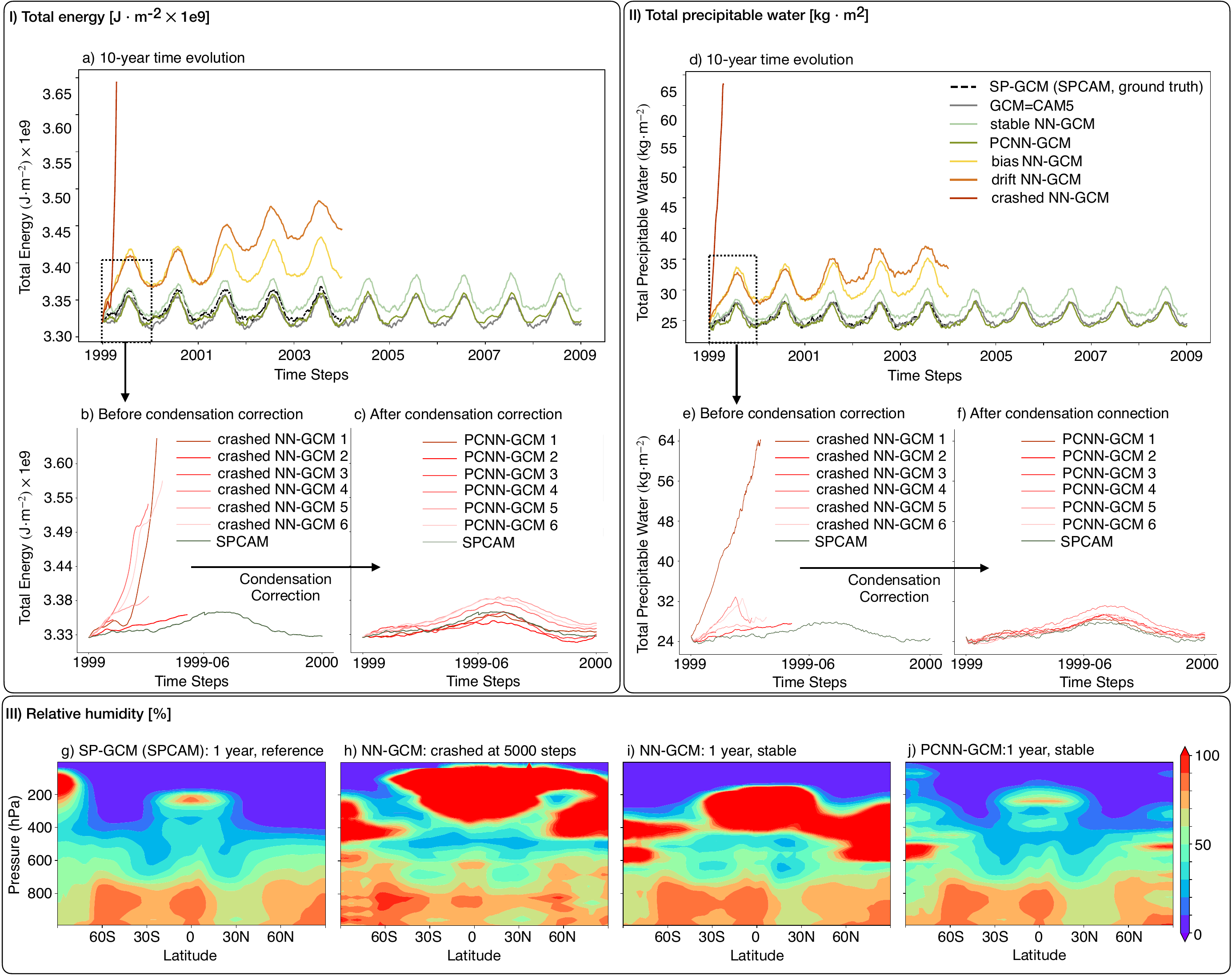}
    \caption{Total energy (a) and total precipitable water (d) time evolution for different models, including stable (light green), biased (yellow), drifted (orange), and crashed (dark red) NN-GCM, PCNN-GCM (dark green), as well as SPCAM (dashed black). 
    Total energy (b) and total precipitable water (e) time evolution of different unstable NN-GCM models. 
    Total energy (c) and total precipitable water (f) time evolution of stable CondensNet models, part of PCNN-GCM, using as baseline the same configuration of the unstable NN-GCM, to show the effects of CondensNet stabilization properties. 
    Relative humidity for SPCAM reference (g), NN-GCM model failing after 5000 time steps (h), stable NN-GCM, and new PCNN-GCM featuring CondensNet (j).}
  \label{fig:Fig2}
\end{figure}

To understand stability in hybrid climate simulations, we use the same GCM configuration as~\cite{wang2022stable}, referred to as NN-GCM (Neural Network-GCM), and reproduce the stable case, the climate drift case, and the crashed cases.
Figure~\ref{fig:Fig2}, panel I, shows the temporal evolution of total energy (total energy refers to the vertically integrated atmospheric total energy -- internal + kinetic + potential -- diagnosed by CAM’s standard energy budget diagnostics, reported in $J \cdot m^{-2}$ and printed in the CESM history/log outputs), while Figure~\ref{fig:Fig2}, panel II, shows the temporal evolution of global average total precipitable water content.
For both panels, dashed black is the SPCAM reference, gray is CAM5, light green is the stable and unbiased NN-GCM, yellow is NN-GCM with bias, orange is NN-GCM causing climate drift, dark red is an unstable NN-GCM that fails on the 107th simulated day (5105th time step), and dark green is the new PCNN-GCM that produces a stable, unbiased behavior, that follow closely the SPCAM reference. 
In simulations that fail or produce climate drift, there is an energy surge (Figure~\ref{fig:Fig2}a), that manifests as an abnormal increase in total precipitable water content (Figure~\ref{fig:Fig2}d). 
Indeed, the behavior of the two variables is qualitatively similar.

Further analysis of the total precipitable water content (i.e., water vapor) distribution in the crashing cases reveals that, as time progresses toward the point of failure, the simulated water vapor shows a significant abnormal increase at 200 hPa and higher vertical levels (Figure~\ref{fig:Fig2}h). 
This suggests that DL parametrizations without physical constraints cause abnormal condensation of water vapor, leading to non-physical relative humidity values.

As water vapor is closely linked to Earth's water cycle and influences the exchange of energy, momentum, and matter among the atmosphere, land, and oceans~\cite{trenberth2009earth}, its unphysical representation can affect several Earth system's processes (and especially the ones related to the water cycle), thereby affecting simulation stability. 
We hypothesize that ensuring a physically accurate representation of condensation is essential for the stability of long-term climate simulations.

To ensure a physical representation of condensation, we introduce an adaptive physical constraint neural network architecture, termed the condensation correction network (ConCorrNet) -- details of ConCorrNet are provided in Methods, section~\ref{sec:methods}.
ConCorrNet is tasked to maintain balance in the water cycle process, and it is integrated with a BasicNet i.e., a neural network architecture that is tasked to predict basic tendencies of water vapor (d$Q$) and dry-static-energy (d$s$). 
ConCorrNet and BasicNet are integrated together to form the new DL parametrization, namely CondensNet, that captures fundamental cloud physics from super parametrization models. 
CondensNet is then integrated into a GCM, namely CAM version 5.2, to form the overall hybrid DL framework, namely PCNN-GCM. The latter is thoroughly described in Methods, section~\ref{sec:methods}.

If we re-run the unstable NN-GCM simulations shown in Figure~\ref{fig:Fig2} (i.e., the crashed NN-GCM simulations depicted in red) using the new PCNN-GCM framework, we now achieve stable simulations without any need for parameter tuning. 
In particular, we tested PCNN-GCM on six unstable NN-GCM models that led the associated simulations to crash~\cite{wang2022stable}. 
Figure~\ref{fig:Fig2}b,e show the total energy and total precipitable water curves before implementing the physical constraint via CondensNet, that is: running the DL parametrization within NN-GCM, which lacks physical constraints.
Figure~\ref{fig:Fig2}c,f show the same curves, using the new PCNN-GCM framework, that imposes a physical constraint on water vapor via CondensNet.
The results show that, after employing CondensNet, the total energy curves of the six previously unstable NN-GCM models align closely with SPCAM (ground truth) and remain stable.
In addition, the water vapor of PCNN-GCM (Figure~\ref{fig:Fig2}j) closely resembles the SPCAM reference (Figure~\ref{fig:Fig2}g), and significantly improves the rather inaccurate results provided by NN-GCM (Figure~\ref{fig:Fig2}i).
In Supplementary Information section~A.5, we show 50-year long simulations, to complement the 10-year ones shown in Figure~\ref{fig:Fig2}.
We remark that the simulations shown are under real-world settings that include land and sea components through CLM version 4.0 and AMIP, respectively.

\subsection{Implications on simulation accuracy}
\label{subsec:accuracy}
The new PCNN-GCM framework, featuring CondensNet, generates stable long-term climate simulations. 
Yet, it requires thorough evaluation to guarantee that the accuracy does not deteriorates due to the novel architecture. 

We compare PCNN-GCM, against NN-GCM, and CAM5, using SPCAM as the ground truth (and training reference for PCNN-GCM and NN-GCM), where the evaluation period spans from 1 January 1999, to 31 December 2003, whereby the training period ranges from 1 January 1997 to 31 December 1998.  
CAM5 is introduced as a widely used baseline GCM, and it represents the host GCM adopted by the PCNN-GCM framework. 
In the results, we refer to CAM5 as GCM=CAM5. 

In Table~\ref{tab:Tab1}, we show the root mean square error (RMSE), calculated as in Equation~\ref{eq:rmse}, between each model and the SPCAM reference, for key simulation variables, noting that before calculating the RMSE we calculated the respective means following Equation~\ref{eq:clim-mean} -- see Methods, section~\ref{sec:means} for more details. 
This comprehensive evaluation quantitatively assesses each model's ability to reproduce SPCAM's climate characteristics across multiple spatial and temporal scales, where SPCAM implements a super parametrization that better represents the under-resolved processes than standard subgrid parametrizations~\cite{kooperman2016robust,arnold2014effects,bretherton2014cloud}.
\begin{table}[htbp] 
  \centering
  \caption{Comprehensive evaluation of climate simulations by SPCAM, CAM5, NN-GCM, and PCNN-GCM. 
  The table shows the reference values from SPCAM and the performance metrics (RMSE) of other models along with their actual values in parentheses. 
  Bold and underlined values indicate the best and second-best performance in comparison metrics.}
  \begin{tabular}{l*{4}{c}}
    \toprule
    & \multicolumn{4}{c}{\textbf{Model Performance}} \\
    \cmidrule{2-5}
    & \multicolumn{1}{c}{SPCAM} & \multicolumn{3}{c}{RMSE (Actual Value)} \\
    \cmidrule{3-5}
    Variables & Actual Value & CAM5 & NN-GCM & PCNN-GCM \\
    \midrule
    Precipitation              & 2.824   & \underline{0.783} &  0.894            & \textbf{0.708}    \\
    (mm/day)                   &         & (2.967)           & (2.835)           & (2.908)           \\[0.5em]
    Precipitation (land)       & 2.198   &  0.761            & \underline{0.719} & \textbf{0.587}    \\
    (mm/day)                   &         & (2.123)           & (2.384)           & (2.251)           \\[0.5em]
    Precipitation (ocean)      & 3.214   & \underline{0.818} &  0.990            & \textbf{0.770}    \\
    (mm/day)                   &         & (3.460)           & (3.146)           & (3.319)           \\[0.5em]
    Total precipitable water   & 25.640  & \textbf{1.358}    &  2.077            & \underline{1.459} \\
    ($kg/m^2$)                 &         & (25.615)          & (26.898)          & (25.554)          \\[0.5em]
    Surface water flux         &  2.825  & \underline{0.342} &  0.382            & \textbf{0.247}    \\
    ($mm/m^2$)                 &         & (2.967)           & (2.724)           & (2.854)           \\[0.5em]
    Sensible heat flux         & 19.493  & \textbf{4.149}    &  6.771            & \underline{4.567} \\
    ($W/m^2$)                  &         & (18.299)          & (20.359)          & (19.973)          \\[0.5em]
    2m temperature             & 287.355 & \textbf{0.622}    &   3.978           & \underline{2.530} \\
    (K)                        &         & (287.121)         & (288.854)         & (288.048)         \\[0.5em]
    10m wind speed             &   6.080 & \textbf{0.434}    &    0.737          & \underline{0.445} \\
    (m/s)                      &         & (5.973)           & (5.625)           & (5.912)           \\[0.5em]
    Zonal-mean specific humidity & -       & 0.231             & 0.144             & \textbf{0.116}    \\
    (g/kg)                       &         & (-)               & (-)               & (-)               \\[0.5em]
    Vertical wind speed        & -       & \textbf{1.894}    & 3.382             & \underline{2.286} \\
    (m/s)                      &         & (-)               & (-)               & (-)               \\[0.5em]
    Zonal-mean temperature     & -       & 2.567             & \textbf{1.419}    & \underline{1.357} \\
    (K)                        &         & (-)               & (-)               & (-)               \\
    \bottomrule
  \end{tabular}
  \label{tab:Tab1}
\end{table}
The results show how PCNN-GCM produces more accurate results against SPCAM than CAM5 and NN-GCM for all variables related to the water cycle (except for total precipitable water, where the results is close to CAM5, yet significantly better than NN-GCM). 
In addition, PCNN-GCM is comparable to CAM5 and significantly better than NN-GCM in terms of sensible heat flux, and wind speed (both 10m and vertical). 
PCNN-GCM performs worse than CAM5 for 2m temperature, yet significantly improving the results of NN-GCM, while it outperforms CAM5 for the vertical distribution of temperature. 

The under-performance in 2m temperature predictions stems from the practical difficulty in accessing boundary layer data within the host GCM (i.e., CAM5.2).
Since the 2m temperature is diagnostically calculated using surface and lowest-level temperatures, an accurate representation of boundary layer processes (surface-atmosphere interactions, turbulent mixing) is crucial. 
Currently, both NN-GCM and PCNN-GCM are not accessing planetary boundary layer variables, such as lowest‑model‑level temperature, surface turbulent fluxes, and a measure of turbulent intensity or mixing, among others. 
This aspect affects the capacity of NN-GCM and PCNN-GCM to capture near-surface temperature gradients.
Accessing these variables would require significant software engineering resources that are beyond the scope of this work.
Yet, we remark that this limitation does not affect the outcomes of this study -- indeed, where boundary-layer information are more easily accessible (e.g., Energy Exascale Earth System Model (E3SM)~\cite{tang2023fully}), these can be retrieved and can provide a more accurate representation of 2m and surface temperature.

\begin{figure}[htbp]
  \centering
  \includegraphics[width=1.0\linewidth]{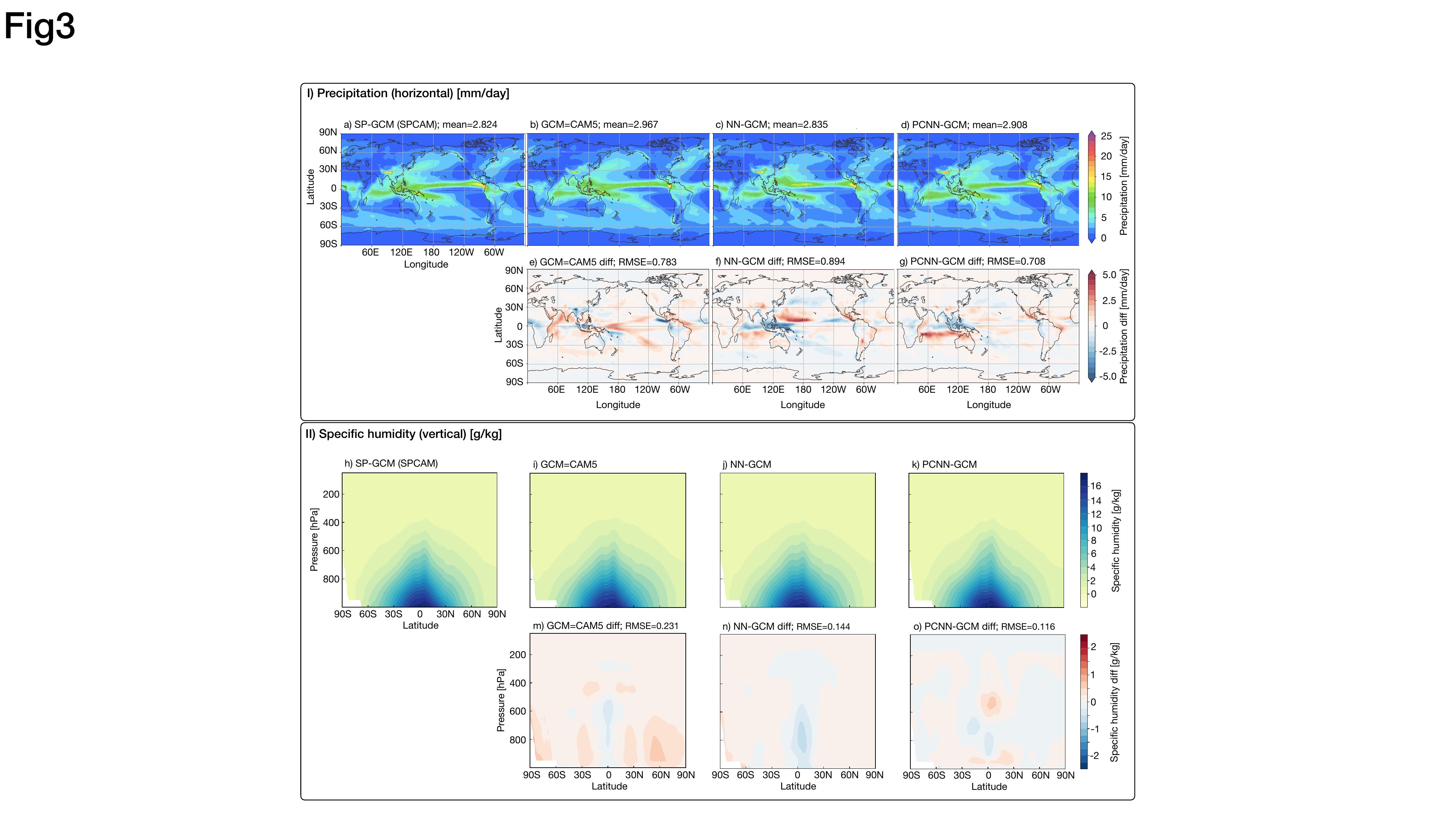}
    \caption{Precipitation (a--d) and zonal-mean specific humidity (latitude--pressure; h--k), for SPCAM, CAM5, NN-GCM, and PCNN-GCM, respectively, and corresponding differences with respect to SPCAM reference (e--g, for precipitation, and m--o, for specific humidity). The fields are annual means (1999-2003) computed as reported in Methods~\ref{sec:means}, and their differences are computed as in Equation~\ref{eq:pattern-diff}, reported in Methods~\ref{sec:errors}. 
    We also provide error metrics, namely RMSE, for all subfigures related to differences (i.e., e--g for precipitation and m--o for specific humidity).}
  \label{fig:Fig3}
\end{figure}

We further assessed the performance of PCNN-GCM for two key variables related to the water cycle, given the physical condensation constraints imposed. 
These are the precipitation field, and the vertical profile of specific humidity in the period 1999--2003, both reported in Figure~\ref{fig:Fig3}. 
For the two variables, we present averaged results, obtained as reported in Methods, section~\ref{sec:means}, where Figure~\ref{fig:Fig3}a--d depict precipitation for the reference SPCAM and for CAM5, NN-GCM, and PCNN-GCM, respectively, whereas Figure~\ref{fig:Fig3}h--k depicts specific humidity.
We also show the corresponding differences of each model (i.e., CAM5, NN-GCM, and PCNN-GCM) with respect to SPCAM, for both precipitation (Figure~\ref{fig:Fig3}e--g) and specific humidity (Figure~\ref{fig:Fig3}m--o), calculated using Equation~\ref{eq:pattern-diff} in Methods (section~\ref{sec:errors}), where we also report the RMSE errors in the title of each subfigure. 
For more details on these calculations, refer to Methods, section~\ref{sec:errors}. 

In terms of precipitation, we observe that all the models considered can reproduce SPCAM results over Asian land, tropical land, and ocean continents. 
However, CAM5 significantly underestimates the precipitation rate in the equatorial eastern Pacific and overestimates the precipitation in tropical continents (Figure \ref{fig:Fig3}e), and NN-GCM underestimates the precipitation in the maritime continent and equatorial eastern Pacific (Figure \ref{fig:Fig3}f). 
Compared to CAM5 and NN-GCM, PCNN-GCM (Figure \ref{fig:Fig3}g) provides results that are significantly closer to SPCAM, with an overall lower RMSE error (see title of each subfigure in Figure~\ref{fig:Fig3}e--g). 
In addition, if we focus on critical regions such as the Intertropical Convergence Zone (ITCZ), the South Pacific Convergence Zone (SPCZ), and the Asian monsoon region, we observe that NN-GCM and PCNN-GCM are closer to SPCAM than CAM5.
Notably, NN-GCM exhibits a precipitation separation in the ITCZ (Figure \ref{fig:Fig3}c), which is absent in PCNN-GCM (Figure \ref{fig:Fig3}d), rendering it closer to the SPCAM distribution.

In terms of specific humidity vertical profiles, Figure~\ref{fig:Fig3}h--k shows that both NN-GCM and PCNN-GCM produce results that are closed to SPCAM than CAM5, as illustrated by the RMSE errors reported in the title of each Figure~\ref{fig:Fig3}m--o.
Owing to CondensNet, that incorporates physical constraints on water vapor condensation, PCNN-GCM (Figure~\ref{fig:Fig3}o) exhibits closer agreement with SPCAM than NN-GCM (Figure~\ref{fig:Fig3}n).
In addition, PCNN-GCM alleviates the dry bias below 600 hPa in the Antarctic region (i.e., 90°S-60°S), leading to better consistency with SPCAM than CAM5 and NN-GCM. 
However, moderate dry and wet biases persist in the tropical region below 400 hPa (Figure~\ref{fig:Fig3}o).

We present additional results in Supplementary Information~A, where we show more detailed analyses of wind speed and temperature, precipitation, and climate variability.
These results confirm that PCNN-GCM better emulates SPCAM compared to NN-GCM across different features, including: annual means (Supplementary Information section~A.1), precipitation distribution (Supplementary Information section~A.2), and climate variability (Supplementary Information section~A.3 and~A.4). 
We further provide a comparison against ERA5 reanalysis data~\cite{hersbach2020era5} and TRMM observational data~\cite{liu2012tropical} (Supplementary Information section~B). 
The results of this comparison show SPCAM (and as a consequence PCNN-GCM) producing results relatively different from ERA5 and TRMM. 
Yet, SPCAM (and PCNN-GCM) shows its strength in capturing tropical wave dynamics as shown by the Wheeler-Kiladis diagrams and the Madden-Julian Oscillation analysis (Supplementary Information sections~B.3 and~B.4). 
SPCAM and PCNN-GCM also performs well in terms of precipitation distribution, being closed to ERA5 and TRMM than e.g., CAM5 (Supplementary Information section~B.2).
Indeed, for both features (tropical wave dynamics and precipitation distribution), SPCAM and PCNN-GCM provide results that are significantly closer to ERA5 and TRMM than CAM5.
These results are expected, as the main focus of this work is on obtaining reliably stable hybrid simulations while remaining close to the accuracy of the super parametrization being emulated (i.e., SPCAM), without applying any bias correction for reanalysis and/or observations.

\subsection{Computational costs}
\label{sec:computational-costs}
The computational experiments were conducted on an HPC cluster composed of eight compute nodes, each equipped with an Intel Xeon Gold 6132 CPU with 24 cores (one MPI process per core) and a single NVIDIA V100 GPU to accelerate the PCNN-GCM computations.
The RAM available was 96GB (128GB/s), with the connection network being Infiniband QDR, and CentOS 7.6.1810 operating system. The compiler adopted was the Intel compiler 19.0.5, and we used the Intel MPI Library 2019 (update 5). 

The training costs of our DL parametrization, namely CondensNet (i.e., the DL parametrization in our PCNN-GCM model) are relatively low: we spent a total of 40 GPU hours for training CondensNet, where we run 50 epochs for BasicNet (29 GPU hours), and 120 for ConCorrNet (11 GPU hours). 
BasicNet is constituted of 7 residual blocks (14 hidden layers) and two separate Residual Multilayer Perceptron (ResMLP) neural networks -- one predicting $\mathrm{d} Q$ and one predicting $\mathrm{d} s$. 
Each ResMLP has 512 neurons per hidden layer and uses ReLU activation functions. 
ConCorrNet is constituted of 6 residual blocks (12 hidden layers) and two separate ResMLP, one -- one predicting $\mathrm{d} Q_{\mathrm{fix}}$ and one predicting $\mathrm{d} s_{\mathrm{fix}}$. 
Each ResMLP has 512 neurons per hidden layer and uses sigmoid activations. 
Indeed, the overall DL parametrization is relatively small, with a total number of parameters equal to 1.7 million (1 million for BasicNet and 700,000 for ConCorrNet).
For further details on the novel DL architecture and on training, the interested reader can refer to Methods section~\ref{sec:methods} and~\ref{sec:methods-training}, respectively. 
\begin{figure}[htbp]
\centering
\includegraphics[width=1.0\linewidth]{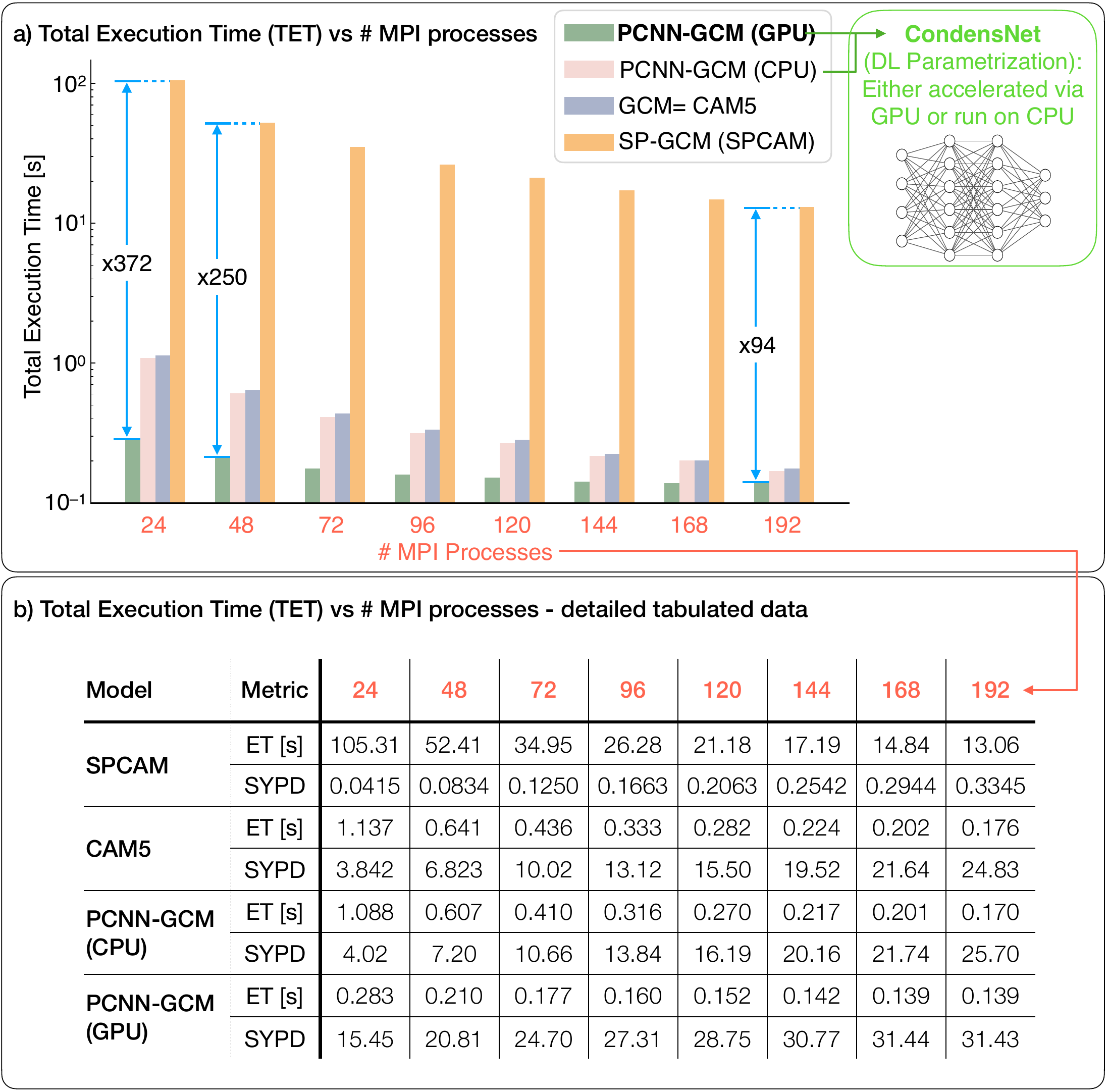}
\caption{Subfigure (a) shows the execution time (ET) in seconds for one simulation time step across different models, namely SPCAM, CAM5, PCNN-GCM, and for different numbers of MPI processes. 
Subfigure (b) provides a detailed view of ET and of the simulated years per day (SYPD) for each model for different MPI processes. 
DL parametrization, CondensNet, inference can be run on both CPU or GPU; we report both results.}
\label{fig:Fig4}
\end{figure}

Figure~\ref{fig:Fig4} shows the execution time of SPCAM (that is the super parametrization reference emulated by CondensNet in our new PCNN-GCM framework), CAM5, and PCNN-GCM. 
We do not report the results for NN-GCM, as they are nearly identical to PCNN-GCM.
The host GCM is run on CPU-only hardware, whereas the DL parametrization inference can be run on either CPU or GPU. 
To this end, CAM5, SPCAM and the host GCM in PCNN-GCM are run on CPU-only hardware, while CondensNet (the DL parametrization in PCNN-GCM) is runnning on CPU (pink color) and GPU (green color).
The results show a significant speed up of PCNN-GCM with respect to SPCAM, when using both CPUs and GPUs for CondensNet (DL parametrization) inference, where inference on CPU is denoted as PCNN-GCM (CPU) and inference on GPU is denoted as PCNN-GCM (GPU), respectively.
In addition, and as one might expect, the GPU-accelerated version of PCNN-GCM is significantly faster than the CPU version, achieving an approximate speedup of 372x relative to SPCAM for 24 MPI processes. 
Increasing MPI process counts enhances CPU-only performance, thereby reducing the relative advantage of PCNN-GCM (GPU), as the DL parametrization remains independent of CPU parallelism. 
Yet, at a moderate concurrency level (e.g., 48 MPI processes), PCNN-GCM (GPU) maintains approximately 250x the performance of SPCAM. 
At 192 MPI processes, while absolute performance improves in all configurations, the relative speedup of PCNN-GCM(GPU) decreases to about 94x as CPU-only models approach near-linear scaling with process counts.

Performance metrics were obtained from CESM’s built-in timing diagnostics in CESM 1.1.1 (the \texttt{ccsm\_timing.\$CASE.\$date} file under \texttt{\$CASEROOT/timing/}), which report simulated years per wall-day (SYPD), PE-hours per simulated year, seconds per model day, and per-component timings; we follow these standard CESM metrics throughout.

These results highlight the potential of GPU-accelerated DL parametrizations to deliver substantial performance gains without the need for large-scale HPC infrastructures. 
Under moderate concurrency, a single workstation equipped with a high-performance GPU can approximate the throughput of much larger CPU-focused deployments -- e.g., running SPCAM for 6 years takes approximately 18 days in a 192 physical CPU cores; while PCNN-GCM takes about 0.191 days (4.6 hours) when accelerated using an NVIDIA Tesla V100, or in roughly 0.233 days (5.6 hours) when running solely on the CPU. 
(see also Figure~\ref{fig:Fig4}).

This shows the extreme potential of our hybrid DL modeling strategy with physical constraints, that can open the path to learning expensive equation-based physics (including for instance large-eddy simulation models) and provide physically-consistent and stable results at a fraction of the computational time.  

For implementation details of the runtime coupling between the Fortran-based host and the Python-based DL parametrization, and for a discussion of scalability limitations and potential remedies, please refer to Supplementary Information section~D.

\section{Discussion}
\label{sec:discussion}
The integration of DL with equation-based models has opened new avenues for addressing long-standing challenges in accurately representing cloud and convection processes, leading to the promising field of hybrid climate modeling. 

Current research has focused on developing DL parametrizations that provide both long-term stability and physical consistency while achieving significantly faster computational performance (e.g., speedup of 10x to 1000x than standard GCM with super parametrization schemes).
Yet, significant breakthroughs are still limited, especially for achieving long-term stability without compromising accuracy.

This work introduces CondensNet, a novel DL parametrization, that embeds physical condensation constraints, to address long-term stability and accuracy issues. 
CondensNet is built on the important finding that the regulation of water vapor saturation is a significant factor affecting simulation stability in real-world land and ocean configurations. 
Unlike traditional super parametrization methods that can suppress anomalous water vapor transport, unconstrained DL parametrizations may fail to replicate moisture regulation, resulting in water vapor oversaturation.  
CondensNet directly learns saturation adjustment processes from super parametrization models (where in this work we used SPCAM), through a Condensation Correction Network (ConCorrNet), that works concurrently and adaptively with a residual multi-layer perceptron, namely BasicNet (see also Methods, section~\ref{sec:methods}). 

The resulting improvement in stability does not compromise predictive accuracy performance or the representation of energy and matter cycles (further results on this aspect are reported in Supplementary Information~C.1).
Moreover, the explicit inclusion of physical constraints enhances interpretability of the hybrid modeling, highlighting the crucial role played by the condensation process. 

CondensNet is integrated in our hybrid DL framework, namely PCNN-GCM, and provides physically-consistent and stable long-term climate simulations, while providing significant speedups compared to the super parametrization benchmark, SPCAM, adopted. 

In particular, we show how PCNN-GCM, through the novel CondensNet architecture, is able to run stable long-term climate simulations, following closely the SPCAM reference (Figure~\ref{fig:Fig2}, in section~\ref{subsec:stablity}), while maintaining physically-consistent results (Table~\ref{tab:Tab1} and Figure~\ref{fig:Fig3}, in section~\ref{subsec:accuracy} as well as Supplementary Information section~A). 
In addition, we show how the PCNN-GCM framework provides speedups in the order of 100x to 372x (depending on the number MPI ranks adopted), compared to the SPCAM reference (Figure~\ref{fig:Fig4}, in section~\ref{sec:computational-costs}). 
PCNN-GCM is also faster and more accurate than CAM5 in certain aspects, such as precipitation and intraseasonal variability, and unlike CAM, supports substantial GPU acceleration. 
For example, when run on modest computational resources (a single 24-core node), PCNN-GCM is nearly four times faster than CAM5 with GPU acceleration.

In comparison to recent approaches that impose humidity constraints or partially address condensation-related issues~\cite{fuchs2023torchclim,behrens2024improving}, CondensNet directly applies adaptive physical constraints to mitigate water vapor oversaturation. 
The work is aligned with several efforts in the community to embed physics in DL models, including the pioneering work on physics-informed neural networks (PINNs)~\cite{raissi2019physics,karniadakis2021physics}.

The new CondensNet DL parametrization embedded in PCNN-GCM sets a milestone on how to implement physical constraints in climate models, that can be readily extended to other atmospheric processes.  
More specifically, CondensNet DL parametrization can emulate other super parametrization models, with e.g., higher resolution, and the overall computational efficiency of PCNN-GCM can be improved using e.g., model compression~\cite{hoefler2021sparsity}, and mixed-precision training~\cite{duben2017study}. 
A similar approach can also be used in recent promising hybrid DL modeling efforts, namely NeuralGCM~\cite{kochkov2024neural}, and purely DL-driven climate models, namely ACE2~\cite{watt2024ace2}, where physical constraints are currently imposed as a post-processing step.

PCNN-GCM marks a significant step towards stable, physically-consistent, and computationally lightweight hybrid climate simulations, that can support better results in regions where climate projection uncertainty still dominates (e.g.,~\cite{dong2024indo}).


\section{Methods}
\label{sec:methods}

\subsection{CondensNet DL architecture and GCM model}
\begin{figure}[htbp]
    \centering
    \includegraphics[width=0.90\linewidth]{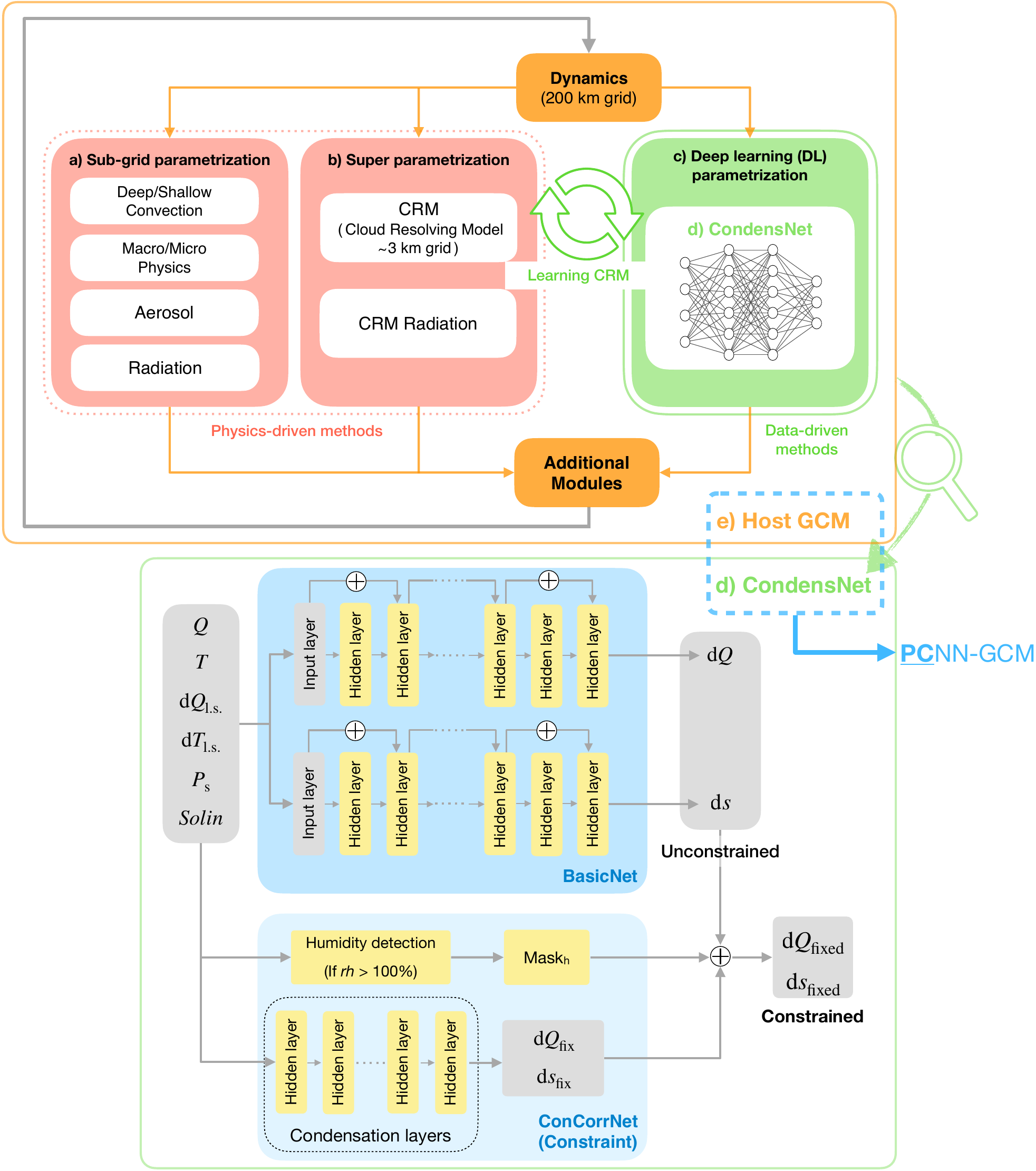}
    \caption{The PCNN-GCM framework. Panels(a) and(b) show the conventional subgrid parametrization and super parametrization approaches, respectively; panel(c) highlights the DL parametrization concept, and panel(d) details the internal architecture of CondensNet; panel(e) is the host GCM. The host GCM plus CondensNet form the PCNN-GCM framework.}
\label{fig:methods}
\end{figure}

CondensNet (Figure~\ref{fig:methods}d) is a novel DL parametrization that learns and emulates the high-resolution cloud-resolving model (CRM) of SPCAM's super parametrization~\cite{khairoutdinov2001cloud} (Figure~\ref{fig:methods}b), where the atmospheric dynamics is driven by the Community Atmosphere Model version 5.2 (CAM5.2)~\cite{neale2010description} (Figure~\ref{fig:methods}e), running at a horizontal resolution of $1.9^\circ \times 2.5^\circ$ with 30 vertical pressure levels, extending up to approximately 2 hPa, and employing a simulation timestep of 30 minutes.
CAM5.2 is further coupled with the Community Land Model version 4.0 (CLM4.0)~\cite{lawrence2011parametrization}, using prescribed sea surface temperatures and sea ice concentrations according to the Atmospheric Model Intercomparison Project (AMIP) protocol. 
Traditional GCMs like CAM use subgrid parametrization based on empirical models to represent cloud and convective processes (Figure~\ref{fig:methods}a), which can introduce significant uncertainties. 
GCMs that use super parametrizations, like SPCAM~\cite{khairoutdinov2001cloud}, mitigate this issue by embedding a high-resolution cloud-resolving model (CRM) within each coarse grid cell (Figure~\ref{fig:methods}b). 
In our study, the two-dimensional CRM of SPCAM consists of 32 grid points in the zonal direction and shares 30 vertical levels with the host model dynamics driven by CAM5.2. 
The host GCM includes all model components except for the parametrizations, namely: the dynamical core, the land model (CLM4.0), and the sea surface temperatures. 
Consequently, SPCAM, CAM5.2, and the hybrid modeling framework share identical host GCM components and simulation data coupling workflows. 

The host GCM provides input variables including large-scale state variables such as water vapor $Q$, temperature $T$, surface pressure $P_{\mathrm{s}}$, and top-of-atmosphere solar insolation $Solin$. 
In addition, large-scale forcing variables such as water vapor forcing $\mathrm{d} Q_{\mathrm{l.s.}}$ and temperature forcing $\mathrm{d} T_{\mathrm{l.s.}}$ are supplied to further enhance the model’s predictive capability. 
CondensNet, our DL parametrization, inherits these input variables and returns predictions of water vapor tendency $\mathrm{d} Q$ and dry-static-energy tendency $\mathrm{d} s$ at each vertical level, 
using an independent ResMLP model from ~\cite{wang2022stable}) to predict downwelling solar radiation fluxes to drive the coupled land surface model.

The complete list of inputs and outputs is provided in Table~\ref{tab:baselinevars}.

Our new DL parametrization, namely the \textbf{CondensNet} model, consists of two neural networks, that have different tasks, and that are integrated together, as depicted in Figure~\ref{fig:methods}d. 
These are:
\begin{itemize}
    \item \textbf{BasicNet}. Tasked to predict basic tendencies of water vapor ($\mathrm{d} Q$) and dry-static-energy ($\mathrm{d} s$), capturing fundamental cloud physics. Here, we use the ResMLP model from ~\cite{wang2022stable}) as a basic model to explore the impact of ConCorrNet on stability in a more intuitive and controllable way.
    
    \item \textbf{Condensation Correction Network (ConCorrNet)}. ConCorrNet is designed to predict physically-constrained corrections to BasicNet's outputs. It is also an NN, operating as an independent, predictive module that is only activated adaptively in regions of unphysical atmospheric conditions (mainly oversaturation) to prevent model instabilities. 
\end{itemize}
CondensNet predicts physically constrained tendencies of water vapor ($\mathrm{d} Q$) and dry-static-energy ($\mathrm{d} s$) that comply with the saturation adjustment mechanism, while the prediction of radiative fluxes remains untouched (i.e., not corrected), as depicted in Figure~\ref{fig:methods}, panel d.

In particular, we identify oversaturated grid points by comparing the prognostic specific humidity ($Q$) with the saturation specific humidity ($Q^\ast$). A grid point is marked for subsequent adaptive physical constraint if it satisfies the condition $Q > Q^\ast$, which is equivalent to a relative humidity exceeding 100\%.
This process results in the creation of a humidity mask, ($\mathrm{Mask}_{\text{h}}$):
\begin{align}
  \mathrm{Mask}_{\text{h}}(\mathrm{lon, lat, lev}) & = 
    \begin{cases} 1, & \text{if} \; Q > Q^\ast (rh > 100\%) \\[-0.1em] 
      0, & \text{otherwise} 
    \end{cases}
\end{align}
where lon, lat, and lev represent the longitude, latitude, and vertical level indices of the grid points. The saturation specific humidity ($Q^\ast$) represents the maximum mass of water vapor that a unit mass of moist air can hold at a given temperature ($T$) and pressure ($p$). It is precisely defined and commonly approximated as:
\begin{align} \label{eq:q_ast}
    Q^\ast = \frac{\epsilon \cdot e^\ast}{p - (1-\epsilon)e^\ast} \approx \frac{0.622e^\ast}{p}
\end{align}
where, $p$ is the local atmospheric pressure, $e^\ast$ is the saturation vapor pressure at temperature $T$ (calculated using a formulation such as the Goff–Gratch equation \cite{goff1946low}), and $\epsilon = R_d/R_v \approx 0.622$ is the ratio of the specific gas constants for dry air ($R_d$) and water vapor ($R_v$).

We use $\mathrm{Mask}_{\text{h}}$ to mark regions where unphysical oversaturation is likely to occur; this marking directs ConCorrNet's attention to these sensitive regions, but does not automatically enforce a correction. 
Instead, ConCorrNet consists of two neural networks that respectively predict the corrective tendencies $\mathrm{d}Q_{\text{fix}}$ and $\mathrm{d}s_{\text{fix}}$. This adaptive methodology allows the model to learn the necessary physical corrections directly from the SPCAM labels for a given atmospheric state, enabling it to reproduce the complex physics of the reference simulation's condensation processes. 

In particular, the correction terms are then applied to the initial tendencies from BasicNet using the humidity mask $\mathrm{Mask}_{\text{h}}$:
\begin{align}
\mathrm{d} Q_{\text{fixed}} & = \mathrm{d} Q - \mathrm{Mask}_{\text{h}} \odot \mathrm{d} Q_{\text{fix}} \label{subeq:dQ} \\
\mathrm{d} s_{\text{fixed}} & = \mathrm{d} s + \mathrm{Mask}_{\text{h}} \odot \mathrm{d} s_{\text{fix}} \label{subeq:ds},
\end{align}
where $\odot$ denotes element-wise multiplication. 
Through this mechanism, the physical constraints learned by ConCorrNet are integrated with the predictions from BasicNet. 
This ensures that the final outputs of the CondensNet model, the tendencies $\mathrm{d}Q_{\text{fixed}}$ and $\mathrm{d}s_{\text{fixed}}$, are physically consistent. 
Following the methodology of Wang et al. (2022)~\cite{wang2022stable}, the surface precipitation rate is then derived by vertically integrating CondensNet's final prediction for the water vapor tendency, $\mathrm{d}Q_{\text{fixed}}$. 
This process provides the necessary moisture source for the land and ocean components of the host GCM, thereby closing the water cycle.

To validate CondensNet and its ConCorrNet's ability to enforce physical constraints and stabilize simulation, we used six ResMLP models from Wang et al. (2022)~\cite{wang2022stable} recorded as causing unstable simulations. 
With the weights of these unstable ResMLP models frozen to act as our BasicNet, we trained only the ConCorrNet module. 
This end-to-end training was guided by a unified loss function on the final, corrected tendencies ($\mathrm{d}Q_{\text{fixed}}$, $\mathrm{d}s_{\text{fixed}}$), ensuring that backpropagated gradients updated only ConCorrNet’s parameters. 
This experimental design isolates and demonstrates the corrective power of our module (i.e., six corrected cases in Subsection~\ref{subsec:stablity}). 
Further training specifications are provided in Subsection~\ref{sec:methods-training}.

Ablation studies presented in Supplementary Information section C.2 further validate that correcting both $\mathrm{d}Q$ and $\mathrm{d}s$ is crucial for simulation stability.

\subsection{Dataset and training details}
\label{sec:methods-training}

CondensNet uses SPCAM simulation data for training. The specific inputs and outputs are listed in Table~\ref{tab:baselinevars}, including 30 vertical levels of specific humidity $Q$, temperature $T$, large-scale water vapor tendency d$Q_{\mathrm{l.s.}}$, large-scale temperature tendency d$T_{\mathrm{l.s.}}$, as well as single-level surface pressure $P_{\mathrm{s}}$ and single-level incoming solar radiation $Solin$. The spatial dimensions of the original SPCAM training data are detailed in Table~\ref{tab:baselinevars}. The original data were generated with a 30-minute time step, same as CAM5.2. Note that CondensNet is trained in a time-independent manner, with samples drawn directly from the SPCAM dataset.

The output variables are the corresponding tendencies of water vapor~$\mathrm{d} Q$ and dry-static-energy~$\mathrm{d} s$ at each vertical level (30 in total), as well as the four radiation fluxes ($SOLS$, $SOLL$, $SOLSD$, and $SOLLD$) in which reach to surface. 

Notably, in CondensNet, following the traditional column-based parametrization design in GCMs, each neural network instance processes a single atmospheric column independently. 
During training, column samples from different spatial locations are randomly shuffled, as the network only needs to learn the vertical physical processes within individual columns. 
When coupled to the host GCM, CondensNet instances operate independently on each column and physics time step. 
The exchange of mass, momentum, and energy between columns is mediated by the model dynamics, represented through the large-scale tendencies (input variables $\mathrm{d} Q_{\mathrm{l.s.}}$ and $\mathrm{d} T_{\mathrm{l.s.}}$). 
This design maintains the intrinsic parallel efficiency of parametrizations while retaining the essential horizontal coupling provided by the dynamics.

The inputs consist of vertical profiles of atmospheric state variables ($Q$, $T$), large-scale tendencies ($\mathrm{d} Q_{\mathrm{l.s.}}$, $\mathrm{d} T_{\mathrm{l.s.}}$), and surface conditions ($P_{\mathrm{s}}$, $Solin$). 
The outputs include physical tendencies of water vapor ($\mathrm{d} Q$) and dry-static-energy ($\mathrm{d} s$) at each vertical level, along with surface radiation fluxes (summarized in Table~\ref{tab:baselinevars}).
\begin{table}[h]
\small
\centering
\caption{Inputs and outputs of the CondensNet DL parametrization.}
\begin{tabular}{lccc}
\toprule
\textbf{Inputs} & Lev & lon & lat \\
\midrule
Specific humidity $Q$ [g/kg] & 30 & 144 & 96 \\
Temperature $T$ [K] & 30 & 144 & 96\\
Large-scale water vapor tendency $\mathrm{d} Q_{\mathrm{l.s.}}$ [g/kg/s] & 30 & 144 & 96\\
Large-scale temperature tendency $\mathrm{d} T_{\mathrm{l.s.}}$ [K/s] & 30 & 144 & 96\\
Surface pressure $P_{\mathrm{s}}$ [Pa] & 1& 144 & 96 \\
Incoming solar radiation $Solin$ [W/m$^\text{2}$] & 1 & 144 & 96\\
\midrule
\textbf{Outputs} &  Lev & lon & lat\\
\midrule
Water vapor tendency $\mathrm{d} Q$ [g/kg/s] & 30  & 144 & 96  \\
dry-static-energy tendency $\mathrm{d} s$ [K/s] & 30  & 144 & 96  \\
Shortwave heating rate $SOLS$ [W/m$^\text{2}$] & 1  & 144 & 96  \\
Longwave heating rate $SOLL$ [W/m$^\text{2}$] & 1  & 144 & 96  \\
Surface downwelling shortwave radiation $SOLSD$ [W/m$^\text{2}$] & 1  & 144 & 96  \\
Surface downwelling longwave radiation $SOLLD$ [W/m$^\text{2}$] & 1 & 144 & 96  \\
\bottomrule
\end{tabular}
\label{tab:baselinevars}
\end{table}

The basic neural networks, BasicNet, is a pre-trained Residual Multilayer Perceptron (ResMLP) that predicts basic tendencies of water vapor ($\mathrm{d} Q$) and dry-static-energy ($\mathrm{d} s$). 
It contains 7 residual blocks (14 hidden layers in total) and 2 separate ResMLP neural networks -- one for predicting $\mathrm{d} Q$ and one for $\mathrm{d} s$. 
Each ResMLP module has 512 neurons for each hidden layer and uses ReLU activation functions.

The condensation correction network (ConCorrNet) is also a ResMLP designed to adjust the predictions of BasicNet to enforce physical constraints related to water vapor saturation.
ConCorrNet architecture includes 6 residual blocks, each containing 2 fully connected layers with 512 neurons, resulting in a total depth of 12 hidden layers. 
We selected the sigmoid activation function based on its superior convergence performance observed in preliminary experiments.

In terms of training, we focus here on the definition and interpretation of the loss functions used to train CondensNet. 
BasicNet and ConCorrNet can be optimized either jointly or in a two-stage scheme in which BasicNet is pretrained and then frozen while training ConCorrNet. 
In this work we adopt the latter for faster convergence; the mathematical objectives below apply to both training protocols.

We minimize a single supervised loss on the final, mask-corrected tendencies -- $\mathrm{d}Q_{\text{fixed}}$ and $\mathrm{d}s_{\text{fixed}}$ -- as defined in Equation~\ref{subeq:dQ} and ~\ref{subeq:ds}, namely:
\begin{equation}
    L_{\text{CondensNet}} 
    = \frac{1}{N} \sum_{i=1}^{N} \Big[ 
        (\mathrm{d}Q_{\text{fixed}, i} - \mathrm{d}Q_{\text{label}, i})^2 
        + (\mathrm{d}s_{\text{fixed}, i} - \mathrm{d}s_{\text{label}, i})^2 
    \Big].
    \label{eq:condensnet_loss}
\end{equation}
\noindent Here, $N$ denotes the number of training samples (grid columns), and $\mathrm{d}Q_{\text{fixed}}$ and $\mathrm{d}s_{\text{fixed}}$ are the masked, physically corrected outputs of CondensNet. 
This formulation of the minimization problem allows the backpropagation of gradients to both BasicNet and ConCorrNet; the humidity mask then determines when and where ConCorrNet is active, and it needs to learn correcting BasicNet unphysical predictions. 

To better understand the training strategy adopted, it is useful to express the minimization problem as the sum of two interpretable terms. 
First, the standard supervised loss for BasicNet
\begin{align}
    L_{\text{BasicNet}} = \frac{1}{N} \sum_{i=1}^{N}
        \Big[ (\mathrm{d}Q_i - \mathrm{d}Q_{\text{label}, i})^2 
        \quad + (\mathrm{d}s_i - \mathrm{d}s_{\text{label}, i})^2 \Big].
    \label{eq:BasicNet_loss}
\end{align}
\noindent Second, the residual-regression objective for ConCorrNet, restricted to the oversaturated region identified by the binary mask $\mathrm{Mask}_{\text{h}}$ (with cardinality $N_m$):
\begin{align}
    L_{\text{ConCorrNet}} &= \frac{1}{N_m} \sum_{i \in \mathrm{Mask}_{\text{h}}} \Big[
        \left(\mathrm{d}Q_{\text{fix}, i} - (\mathrm{d}Q_i - \mathrm{d}Q_{\text{label}, i})\right)^2 \nonumber \\
        &\quad + \left(\mathrm{d}s_{\text{fix}, i} - (\mathrm{d}s_{\text{label}, i} - \mathrm{d}s_i)\right)^2 \Big].
    \label{eq:ConCorrNet_loss_conceptual}
\end{align}
\noindent Intuitively, if we interpret the humidity mask as a binary gate: when $\mathrm{Mask}_{\text{h}}=1$ ConCorrNet learns to correct the BasicNet predictions, while when $\mathrm{Mask}_{\text{h}}=0$ ConCorrNet is inactive. 
Accordingly, the implemented loss in Equation~\ref{eq:condensnet_loss} back-propagates gradients to both heads wherever $\mathrm{Mask}_{\text{h}}=1$, and to BasicNet everywhere. Details are provided in Supplementary Information section~C.3.

The specific hyperparameters used during training for the results presented in this work are listed in Table~\ref{tab:hyperparameter-fine-tune}, for both BasicNet and ConCorrNet.
\begin{table}[h]
\centering
\caption{Hyperparameter settings for training CondensNet (BasicNet + ConCorrNet).}
\begin{tabular}{lll}
\toprule
Hyperparameter                & BasicNet         & ConCorrNet       \\
\midrule
Activation function           & Relu             & Sigmoid          \\
Training epochs               & 50               & 120              \\
Number of residual blocks     & 7                & 6                \\
Learning rate (initial)       & 0.001            & 0.00075          \\
Learning rate schedule        & Cosine annealing & Cosine annealing \\
Hidden layer size             & 512 neurons      & 512 neurons      \\
Optimizer                     & Adam             & SGD              \\
Batch size                    & 1024             & 768              \\
\bottomrule
\end{tabular}
\label{tab:hyperparameter-fine-tune}
\end{table}
The model was implemented using PyTorch and trained on multiple GPUs to accelerate computation. 
We used standard techniques such as data normalization and weight initialization to enhance training stability. 
Early stopping and model checkpointing were employed to prevent overfitting.
The code is feely available at~\href{https://github.com/MathEXLab/PCNN-GCM}{https://github.com/MathEXLab/PCNN-GCM}.

\subsection{Climatological Processing}
\label{sec:means}

For variables with vertical distribution (temperature, wind speed, specific humidity), the zonal mean at each pressure level is given by
\begin{equation}
    \overline{X}_{\mathrm{zonal}}(\phi,p) \;=\; \frac{1}{N_{\lambda}} \sum_{i=1}^{N_{\lambda}} X(\lambda_i,\phi,p),
     \label{eq:zonal-mean}
\end{equation}
where $N_{\lambda}$ is the number of longitudinal grid points, $\lambda_i$ is the longitude at grid point $i$, $\phi$ is latitude, and $p$ is pressure level. 
For surface or near-surface variables (precipitation, 10m wind speed), the horizontal mean is given by
\begin{equation}
    \overline{Y}_{\mathrm{horizontal}} \;=\; \frac{1}{N_{\lambda}N_{\phi}} \sum_{j=1}^{N_{\phi}}\sum_{i=1}^{N_{\lambda}} Y(\lambda_i,\phi_j)w(\phi_j),
    \label{eq:horizontal-mean}
\end{equation}
where $w(\phi_j)$ is the latitudinal weight factor. 
The climatological means are then obtained by averaging these spatial means over the analysis period
\begin{equation}
    \overline{X}_{\mathrm{clim}} \;=\; \frac{1}{Y}\sum_{y=1}^Y X_{m,y},
    \label{eq:clim-mean}
\end{equation}
where $Y$ is the total number of years in the analysis period, $X_{m,y}$ represents the monthly mean for month $m$ in year $y$.

\subsection{Error Metrics}
\label{sec:errors}

Once the means introduced in section~\ref{sec:means} are obtained, we use different error metrics to asses the performance of PCNN-GCM against NN-GCM and CAM5, using as a reference (i.e., ground truth) SPCAM. In particular, we use the \textbf{pattern difference}
\begin{equation}
    \mathrm{diff}(\phi,\lambda,p) = X_{\mathrm{model}}(\phi,\lambda,p) - X_{\mathrm{SPCAM}}(\phi,\lambda,p)
    \label{eq:pattern-diff}
\end{equation}
where $X_{\mathrm{model}}$ and $X_{\mathrm{SPCAM}}$ represent the climatological means from a given model and SPCAM respectively, the \textbf{weighted root mean squared error} for variables with vertical distribution
\begin{equation}
    \mathrm{RMSE}(p) \;=\; \sqrt{\frac{\sum_{j=1}^{N_{\phi}}\sum_{i=1}^{N_{\lambda}} [X_1(\lambda_i,\phi_j,p) - X_2(\lambda_i,\phi_j,p)]^2 w(\phi_j)}{\sum_{j=1}^{N_{\phi}}w(\phi_j)}}
\label{eq:rmse}
\end{equation}
where $X_1$ and $X_2$ represent the climatological means from two different models (for surface variables, the same formula applies without the pressure level dependency), and the \textbf{coefficient of determination}
%
\begin{equation}
    R^2 = 1 - \frac{\sum_{i=1}^N \bigl(X_i^{\mathrm{model}} - X_i^{\mathrm{SPCAM}}\bigr)^2}{\sum_{i=1}^N \bigl(X_i^{\mathrm{SPCAM}} - \overline{X}_{\mathrm{SPCAM}}\bigr)^2}
\label{eq:r2}
\end{equation}
where $N$ is the total number of samples, $X_i^{\mathrm{model}}$ and $X_i^{\mathrm{SPCAM}}$ are the values at sample point $i$ for a given model and SPCAM respectively, and $\overline{X}_{\mathrm{SPCAM}}$ is the mean of SPCAM values over all samples.

In Equations~\eqref{eq:pattern-diff}, \eqref{eq:rmse}, and \eqref{eq:r2}, $X_{\mathrm{model}}$ correspond to the model being evaluated (i.e., PCNN-GCM, NN-GCM, and CAM5), while $X_{\mathrm{SPCAM}}$ corresponds to the SPCAM reference (i.e., ground truth).

\clearpage

\backmatter

\bmhead{Published version}
The published version of this article (including the supplementary material) can be found on npj Climate and Atmospheric Science website at \url{https://www.nature.com/articles/s41612-025-01269-5}

\bmhead{Data availability}
All data supporting the findings of this study are available at Hugging Face under the persistent identifier \url{https://huggingface.co/datasets/xin-w24/CondensNet}. 
Additional materials are provided in the Supplementary Information.

\bmhead{Code availability}
All custom code used in this study is available at \url{https://github.com/MathEXLab/PCNN-GCM}. 
The code is distributed under the GPL-3.0 license.

\bmhead{Acknowledgements}
W.X.\ acknowledges support from the National Natural Science Foundation of China (Grant U2242210).
G.M.\ acknowledges support from MOE Tier 2: Prediction-to-Mitigation with Digital Twins of the Earth’s Weather, under grant number \# T2EP50221-0017. 

\bmhead{Author contributions}
X.W.\ developed the idea, implemented the methodology, ran experiments and analyses, performed results post-processing and visualization, and drafted the paper.
J.C.\ contributed to results post-processing and visualization, and provided inputs on the paper draft.
J.Y., J.A., S.S.\ provided inputs on computational performance .
K.F., C.C.\ provided inputs on atmospheric science.
T.A., R.M.\ provided inputs on climate modeling.
W.X.\ provided inputs on the paper draft.
G.M.\ supervised and coordinated the work, developed the idea, provided inputs on computational modeling and AI, performed results post-processing and visualization, and wrote the paper.
All authors read and approved the manuscript.

\bmhead{Competing interests}
The authors declare no competing interests.

\clearpage

\bibliography{references}

\end{document}